# Plasmon Injection to Compensate and Control Losses in Negative Index Metamaterials

Mehdi Sadatgol[1], Şahin K. Özdemir[2], Lan Yang[2], and Durdu Ö. Güney[1,*]

[1]*Department of Electrical and Computer Engineering, Michigan Technological University, Houghton, MI 49931, USA*
[2]*Department of Electrical and Systems Engineering, Washington University, St. Louis, MO 63130, USA*

Metamaterials have introduced a whole new world of unusual materials with functionalities that cannot be attained in naturally occurring material systems by mimicking and controlling the natural phenomena at subwavelength scales. However, the inherent absorption losses pose fundamental challenge to the most fascinating applications of metamaterials. Based on a novel plasmon injection (PI or Π)-scheme, we propose a coherent optical amplification technique to compensate losses in metamaterials. Although the proof of concept device here operates under normal incidence only, our proposed scheme can be generalized to arbitrary form of incident waves. The Π-scheme is fundamentally different than major optical amplification schemes. It does not require gain medium, interaction with phonons, or any nonlinear medium. The Π-scheme allows for loss-free metamaterials. It is ideally suited for mitigating losses in metamaterials operating in the visible spectrum and is scalable to other optical frequencies. These findings open the possibility of reviving the early dreams of making "magical" metamaterials from scratch.

PACS numbers:

Metamaterials have led to previously unthought-of applications such as flat lens [1], perfect lens [2], hyperlens [3-5], ultimate illusion optics [6-8], perfect absorber [9,10], optical analog simulators [11,12], metaspacers [13], and many others. Despite tremendous progress in theory and experimental realizations, the major current challenges seem to further delay the metamaterial era to come. For instance, achieving full isotropy, feasible fabrication methods, broad bandwidth, and compensation of dissipative losses especially at optical frequencies are among the challenges yet to be solved [14,15]. Perhaps, the most critical of all is how to avoid optical losses – especially in large-volume structures. The performance of devices utilizing metamaterials dramatically degrades at optical frequencies due to significant ohmic losses arising from the metallic constituents. The strategies proposed to mitigate the losses include passive reduction and active compensation schemes. Avoiding sharp edges [16], reducing skin depths [16-18], and classical analog of electromagnetically induced transparency are proposed as passive loss minimization techniques [19,20]. Constitutive materials such as dielectrics [21], nitrides [22], oxides [23], graphene [24], and superconductors [24-30] have also been explored for possible alternatives to lossy conductors. However, none of these materials have so far shown to outperform the performance of high-conductivity metals at room temperature [31,32]. Active compensation of losses using gain medium has been emerged as the most promising strategy to avoid the deleterious impacts of losses on metamaterial devices. Optically pumped semiconductor quantum dots and quantum wells incorporated in planar metamaterials have been experimentally shown to compensate the losses to some extent [33-38]. A remarkable improvement in the figure of merit (FOM) of a negative refractive index of a fishnet metamaterial operating in the visible range has been recorded using dye molecules embedded in epoxy resin and pumped by picosecond pulses [35,37]. Although important progress has been made in theory [14,15,34,35,38-40] and experiments [14,15,33,36,37,41-43], there still exist concerns about the viability of active compensation of loss by gain medium for practically large-volume metamaterials [14]. First, it was shown that causality and the stability of the system make the loss compensation difficult without compromising interesting properties of the metamaterials [14,44,45]. Second, requirement for optical or electrical pumping makes the metamaterials complex and dependent on available pump sources or gain materials at particular wavelengths. Third, high maintenance and short chemical life of the dye molecules make the durable metamaterials unfeasible [35,37].

Here, we propose a novel optical amplification technique to compensate and control losses in metamaterials. Our proposed amplification scheme is ideally suited for compensating losses in metamaterials operating in the visible spectrum and fundamentally different than major amplification schemes based on stimulated optical emission, Raman scattering, and optical parametric amplification, or recently proposed surface plasmon polariton amplification for plasmonic integrated circuits [43]. It does not require gain medium, interaction with phonons, or any nonlinear medium, and it operates at room temperature. Our technique relies on the constructive interference of externally-injected surface plasmon polaritons (SPPs) within the metamaterial to coherently amplify the domestic SPPs of the metamaterial. The amplified SPPs are then coupled into the free-space via an appropriately designed grating network at the output port of the metamaterial. In this Letter, we show that this proposed scheme, referred to as Π-scheme, provides diverging FOM, hence loss-free optical metamaterials without using optical gain providing medium. Although the negative index metamaterial (NIM) described here is functional for normally incident plane waves only, the Π-scheme can be generalized to arbitrary form of incident waves.

Below, we first describe how the Π-scheme improves the FOM of NIMs based on a conceptual plasmonic structure. Then, we examine the simulations and effective parameter retrieval results [46-49] to characterize and validate the improved performance offered by the proposed scheme.

Fig. 1 illustrates the conceptual structure that is built by arranging the unit cell of a so-called "surface plasmon driven (SPD) NIM" [49,50] into three superlattices. Each superlattice is built by repeating the unit cell, indicated by the red rectangular box, an integer number of times except that the superlattices at the sides have missing grating above the thin film. This prevents the coupling of SPPs back to free-space photons, because the gratings are designed to facilitate the coupling between the photons and the SPPs. The central superlattice is intended as the metamaterial where the losses are to be compensated. As shown in Fig. 1 the SPPs which are excited by the auxiliary ports, $P_3$ and $P_4$, at the side superlattices propagate toward the central superlattice to form a standing wave. Then these SPPs injected from the side superlattices to the central superlattice amplify the domestic SPPs excited by the input port $P_1$ (i.e., no input field at $P_1$ is supplied in Fig. 1 to clearly show injected SPPs only). We design the structure such that all of these three SPP waves interfere constructively at the central superlattice. The amplified SPPs are finally converted back to the free-space photons at the output port $P_2$ facilitated by the grating above the thin film in the central superlattice. Detailed working principle and general theory of the proposed loss compensation scheme and amplification are given in [51].

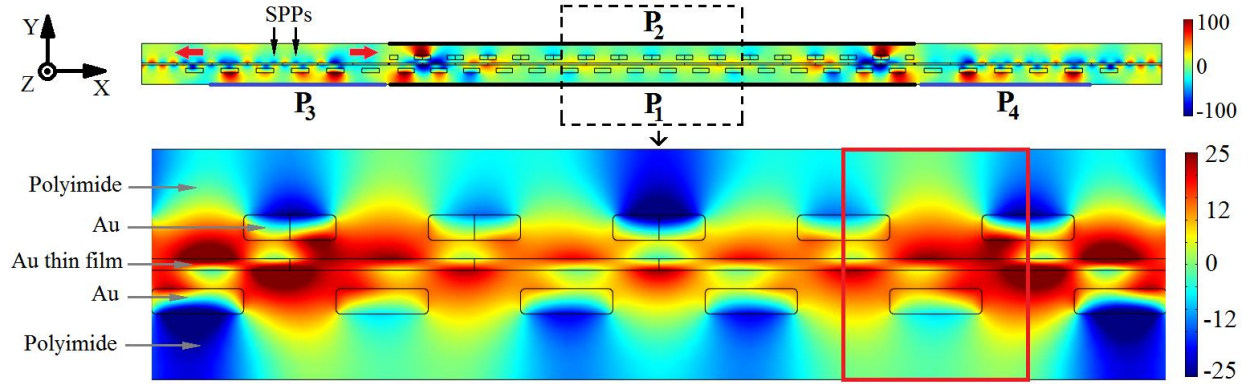

Figure 1: **Conceptual structure for the Π-scheme:** Central superlattice constitutes the metamaterial. $P_1$ and $P_2$ are the input and output ports of the metamaterial, respectively. $P_3$ and $P_4$ are the auxiliary ports through which surface plasmons for amplification are excited externally. The red rectangular box illustrates the metamaterial unit cell consisting of gold grating and the gold thin film (in the center of the unit cell) embedded in polyimide. The thickness of the gold thin film and its distance from the strips are 5 nm and 8 nm, respectively. The thickness of the gold strips in the gratings is 11 nm. Grating period at the central superlattice (i.e., between the ports $P_1$ and $P_2$) is 80 nm and 86 nm at the side superlattices above the ports $P_3$ and $P_4$. There exist 6 periods in each of the side superlattices and 16 periods in the central superlattice. The structure is translationally invariant in the z-direction. SPPs propagate from the side superlattices to central superlattice and excite the metamaterial eigenmode (lower panel), which in turn amplifies the domestic SPPs in the central superlattice if an input field at port $P_1$ exists. Finally, the amplified SPPs couple to free-space modes. The surface plot corresponds to the z-component of the magnetic field (A/m) at magnetic resonance. The middle part of the central superlattice, indicated by a box, is enlarged and the field amplitude is multiplied by 4 to clearly illustrate the mode profile.

We performed frequency domain analysis of the structure using the finite element method in COMSOL Multiphysics to find the multi-port scattering parameters. Drude model with plasma frequency $f_p = 2175$ THz and collision frequency $f_c = 6.5$ THz was used to describe gold [16]. Polyimide layers have relative permittivity $\varepsilon_r = 3.6$. Since the ports $P_3$ and $P_4$ of the structure in Fig. 1 are the auxiliary ports to compensate the dissipated power, the structure in Fig. 1 can be effectively modeled as a 1-input 1-output structure with $P_1$ and $P_2$ being the actual input and output ports, respectively, for the metamaterial (i.e., central superlattice). Thus having in hand the scattering parameters from COMSOL simulation, we calculated the effective refractive index and the wave impedance of our plasmonic structure using the retrieval procedure [17,47-49,58-62] which includes equations based on transformation matrix of a 1-input 1-output homogenous slab. Then the overall scattering parameters of our central superlattice is given as [51],

$$S_{11} = S'_{11} + \sqrt{\alpha/2}(S'_{13} + S'_{14}) \quad \text{(1-a)}$$
$$S_{21} = S'_{21} + \sqrt{\alpha/2}(S'_{23} + S'_{24}) \quad \text{(1-b)}$$

which account for the external contribution of auxiliary ports $P_3$ and $P_4$ to the metamaterial ports $P_1$ and $P_2$. The primed parameters were found from the COMSOL simulation, and the unprimed parameters are the overall parameters corresponding to the two port representation. The coefficient $\alpha/2$ is defined as the ratio of the optical power applied to each of the auxiliary ports $P_3$ and $P_4$ to the optical power applied to the input port $P_1$. The fields applied to the auxiliary ports $P_3$ and $P_4$ are always in phase and have equal power. Although the retrieval procedure assumes an infinite structure along the direction perpendicular to the direction of propagation of the incident field, we have verified that the retrieval procedure is applicable to finite-size ports (Fig. 1) excited by plane waves. This is similar to experimental characterization of finite metamaterial structures with finite beam spot size.

The commonly used FOM to quantify the loss compensation in NIMs is defined as FOM $= -n'/n''$ [34,35,37], which is the same expression as the FOM used to measure the transparency of passive NIMs [15,63,64]. Although we can define the FOM for any arbitrary value of the refractive index, the FOM has a special importance when $n' = -1$ due to the Pendry's perfect lens [2]. We denote this value of the FOM as FOM$_{-1}$. In order to maximize the FOM$_{-1}$ we need to aim for vanishing $n''$ while keeping $n' = -1$. This can be achieved by maximizing the amplitudes $|S'_{23}| = |S'_{24}| = |S'_{2a}|$ and removing the phase shift between $S'_{23}$, $S'_{24}$, and $S'_{21}$ at the frequency where $n' = -1$; because the imaginary part of refractive index, $n''$, is responsible for the power loss in the structure and it can be minimized by increasing the transmittance while preserving the relation [34,38] $A + T + R = 1$ at the central superlattice. Here, absorbance $A = 1 - |S_{11}|^2 - |S_{21}|^2$, transmittance $T = |S_{21}|^2$, and reflectance $R = |S_{11}|^2$. Note that absorbance $A$ of the central superlattice can be negative, hence implying gain.

The simplest design for loss compensated metamaterial can be obtained by using the same unit cell geometry of the original SPD NIM [49] for both the central superlattice and the symmetrically placed auxiliary superlattices with the exception that the gratings above the thin film in the auxiliary superlattices are removed to prevent the coupling of SPPs to free-space photons. Then, by tuning the $\alpha$ and the phase difference between the auxiliary ports and the input port $P_1$, $n''$ can be arbitrarily reduced at $n' = -1$, made zero or even negative. The system does not become unstable and oscillatory when $n''$ becomes negative because the proposed scheme utilizes an open loop system which allows for only limited gain [65]. When $n'' = 0$ is reached at $n' = -1$, the frequency where this occurs may slightly shift [51]. If desired, this shift can be recalibrated through several iterations by slightly adjusting the system parameters. In this process $|S'_{2a}|$ is preferred to be sufficiently large, since larger $|S'_{2a}|$ indicates larger plasmon injection rate, hence smaller required $\alpha$ (i.e., which means better performance). $S'_{2a}$ can be tuned by changing the grating period and the distance between the side and central superlattices.

Below we use an optimized structure giving a maximum value of $|S'_{2a}|$ close to 0.2. The side superlattices of the structure have the grating period of 86 nm and are adjacent to the central superlattice. There exist 6 periods in each of the side superlattices and 8 periods in the central superlattice. The phase shift between the auxiliary ports and the input port $P_1$ is almost zero.

In Fig. 2a we give the retrieved $n'$ and $n''$ for the central superlattice for $\alpha = 6$. It is clearly seen that $n''$ is very close to zero in the negative index region implying low-loss operation. This is attributed to the compensation of the loss at the central superlattice by the SPPs excited through ports $P_3$ and $P_4$. Thus, the central superlattice forms our new low-loss NIM. We estimate the FOM-1 as 52 which is 21 times larger than the FOM-1 achieved by the same structure without the proposed loss compensation [49]. The result for retrieved impedance $z = z' + iz''$ is shown in Fig. 2b. The features observed in the retrieved parameters are qualitatively similar to passive metamaterials including SPD NIMs [49] with the exception of relatively smoother variations since such metamaterials are not as complex as the studied loss compensated metamaterial structure, which involves, for example, finite-size effects and plasmon injection rate varying with frequency [51]. Detailed discussion and subtle points behind the calculations of retrieved optical parameters in Fig. 2 are given in [51].

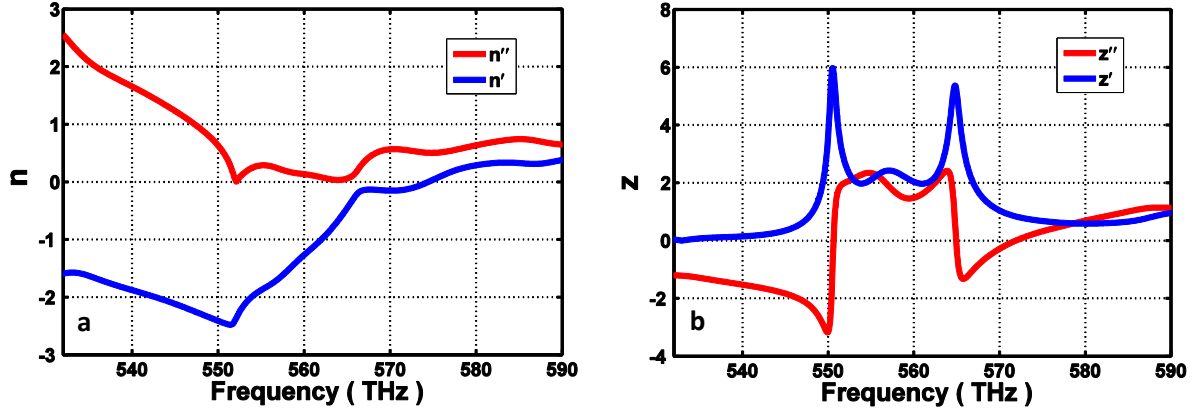

Figure 2: **Effective parameters for a single functional layer at $\alpha = 6$:** (a) Real ($n'$) and imaginary ($n''$) parts of the retrieved effective refractive index. Around 552THz and 562THz, the system behaves as an ultra-low loss NIM. (b) Real ($z'$) and imaginary ($z''$) parts of the impedance.

Fig. 3 shows the overall transmittance $|S_{21}|^2$ through the central superlattice as a function of frequency for different values of $\alpha$. Note that the transmittance increases with increasing $\alpha$ over the most of the spectrum including the negative index region at around 560THz. These results depend on the amount of transferred power into the central superlattice and the phase necessary for the desired loss compensation but not directly on the size of the auxiliary ports [51]. Detailed characterization of the device is given in [51]. We should note here that the output field at port $P_2$ is not just the simple summation of the output powers from individual sources (i.e., $P_1$, $P_3$, and $P_4$) but their coherent summation which amplifies the signals due to constructive interference. Additional details on coherent loss-compensation and signal-amplification characteristics of the device based on phase dependent transmittance are given in [51]. Also there will be leakage to the output port from the auxiliary plasmonic fields even when the input $P_1$ is zero but this is usually very low [51].

We have repeated the simulations for different values of the parameter $\alpha$ and calculated FOM-1. The results are depicted in Fig. 4. FOM-1 rapidly increases with $\alpha$ and starts to diverge at a small value of $\alpha = 7$. The frequency corresponding to FOM-1 changes only slightly around 560THz. In this regime, the medium becomes lossless (i.e., $n''$ approaches to zero). Fig. 4 shows that FOM-1 already becomes substantially large when $\alpha$ is between 6.5 and 7. Therefore, as long as $\alpha$ is tuned to this range, resultant FOM-1 can be regarded as practically diverging which means that the medium becomes practically loss-free. This also makes the implementation experimentally feasible. Thus, the demonstrated rapid growth in FOM-1 shows that our loss compensation scheme is a very effective solution to mitigate any type of losses (i.e., metallic, dielectric, and even losses due to fabrication imperfections) in metamaterials. Furthermore, ultra-low-loss bandwidth around refractive index $n = -1$ can be also broadened [51].

Although the Drude model that we have used for gold underestimates the losses with respect to experimental values, this was chosen only for simplicity and does not invalidate the applicability of our loss compensation scheme to highly lossy systems. Given the rapid growth in FOM-1 with increasing $\alpha$ as shown in Fig. 4 [51], the realistic experimental values for the permittivity of gold would only require larger $\alpha$ for loss-free operation, but not much larger than 7, because the metamaterial rapidly enters the amplifying regime beyond $\alpha = 7$.

The proof of principle plasmonic structures such as that given in Fig. 1 can be fabricated by self-aligned electron-beam lithography exposures and dielectric layer deposition via atomic layer deposition by choosing, for example, Ag as the metal and $Al_2O_3$ as the dielectric. Considering the experimental data [66] and misalignments, we have previously verified [67] that the negative index persists. We have confirmed through a series of sensitivity analyses that the structure is also tolerant to other imperfections possible at different stages of the fabrication process including the variations in dielectric thickness (i.e., change from 8nm to 15nm), dielectric surface roughness (i.e., within ± 10% of the average thickness), and strip width (i.e., ± 2.5% change). Additional fabrication imperfections may further deteriorate the negative index. However, we should note that as long as the underlying eigenmode (see the lower panel in Fig. 1) [51] of the metamaterial survives, the Π-scheme cannot only restore the negative index but also makes the metamaterial loss-free at only a larger $\alpha$ value than predicted theoretically. Similar metamaterial structures employing plasmonic modes, waveguide modes, or surface modes can also be designed and fabricated at lower frequencies with less difficulty. Loss compensation schemes analogous to Π-scheme can be easily devised for other available metamaterials.

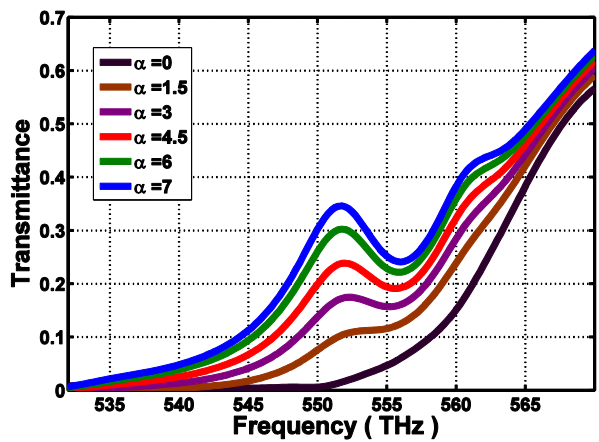

Figure 3: **Transmission spectra of the central superlattice:** Transmittance through central superlattice increases with $\alpha$. The increase in the transmittance is significant in the negative refractive index region around 560THz.

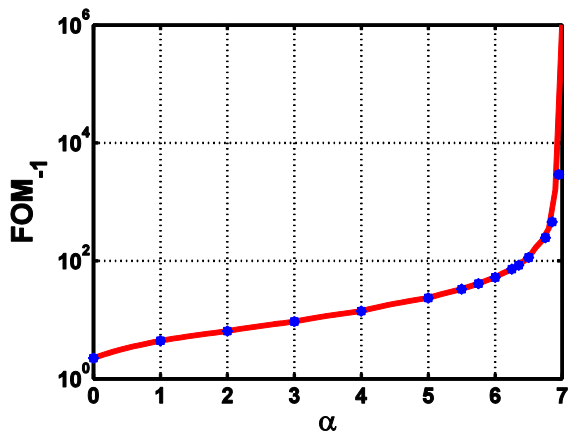

Figure 4: **The figure-of-merit for the negative-index material (FOM$_{-1}$) as a function of $\alpha$:** The blue data points show the FOM$_{-1}$ for the simulated values of $\alpha$. The red solid line is the best-fitting curve. FOM$_{-1}$ diverges at $\alpha = 7$ and the metamaterial becomes loss free. Beyond $\alpha = 7$, the metamaterial becomes an amplifying medium.

We should note that here we considered normal incidence only and accordingly designed the NIM structure. However, the underlying idea of plasmon injection for compensating and controlling losses can be generalized to arbitrary form of incident waves. It is important to point out that the device does not have to know the form of incident waves, because the excitation of required eigenmodes, bulk and surface waves, amplitude and phase relations can be automatically achieved [51]. The greatest challenge that hinders the advance of metamaterials and plasmonics, and preventing real-life applications in particular hyperlenses and superlenses is the loss problem. Our loss compensation scheme may revive these early dreams of the metamaterials. For example, hyperlenses or superlenses with resolution significantly beyond diffraction limit may be possible. Epsilon-near-zero materials [68] and transformation optics [6,7] can also benefit from this method. Our proposed scheme can also be applied to acoustic metamaterials to compensate damping in the same way as it is applied to electromagnetic waves here.

This work was supported by the National Science Foundation under grant ECCS-1202443. We would like to thank Philip G. Evans at Oak Ridge National Laboratory for fruitful discussion on fabrication of proposed SPD plasmonic metamaterial structures, and Jae Yong Suh at Northwestern University on optical gain and amplification in plasmonic structures.

* Corresponding author. dguney@mtu.edu

*Supplementary Material for*

**Plasmon Injection to Compensate and Control Losses in Negative Index Metamaterials**

Mehdi Sadatgol[1], Şahin K. Özdemir[2], Lan Yang[2], and Durdu Ö. Güney[1,*]

[1]*Department of Electrical and Computer Engineering, Michigan Technological University, Houghton, MI 49931, USA*
[2]*Department of Electrical and Systems Engineering, Washington University, St. Louis, MO 63130, USA*

## 1. Theory of Plasmon Injection Scheme for Compensation of Losses in Metamaterials

### 1.1. Introduction

In 2011, we introduced surface plasmon driven metamaterials [1, 2]. Here we show that such metamaterials offer a novel loss compensation scheme due to their unique multiple channel plasmon injection feature as described in detail below.

Fig. S1 shows the surface plot for the *z*-component of the magnetic field distribution inside a surface plasmon driven negative index metamaterial (NIM) [1] at the magnetic resonance frequency of 594THz. This type of NIM structure underlies our loss compensation scheme and corresponds to the central superlattice in Fig. 1. Only a finite number of periods of the infinite structure in the lateral direction was shown in Fig. S1. The field distribution was obtained from wave propagation analysis in frequency domain using finite element based commercial software package COMSOL. The magnetic dipoles contributing to the negative index are generated by the loop currents indicated by dashed circles in Fig. S1. These loop currents are similar to well-known multiple-gap split ring resonator (SRR) [3, 4] loop currents. As can be seen in Fig. S1a, the arrows that represent the electric current density make several rotations. We have approximated the current density with these current loops not only because there are rotations in the current density, but also because these loops approximate the magnetic field very well. Since each current loop is surrounded by three other loops with opposite sense of rotation, magnetic fields add destructively and almost cancel each other outside of the loops. However, there is a net non-zero magnetic field in the unit cell of the structure (i.e., magnetic fields generated by smaller loops cannot completely cancel the magnetic fields generated by larger loops). Fig. S1b shows a simple sketch of the strong magnetic field regions due to the induced current loops. Note that this sketch closely represents the actual magnetic field distribution in Fig. S1a. On the other hand, Fig. S1c shows the equivalent RLC circuits corresponding to current loops for a smaller segment of the structure. Here, the metallic parts contribute to the inductance, while the surrounding dielectric material and gaps between the metals contribute to the capacitance similar to SRRs. These observations self-consistently explain the origin of the magnetic resonance mode. Other details including the design guidelines and retrieved effective parameters for this type of NIM underlying our proposed loss compensation scheme can be found in our previous work [1]. Here, we will rather focus specifically on the working principle and design guidelines for the proposed loss compensation mechanism which is the main scope of the present Letter.

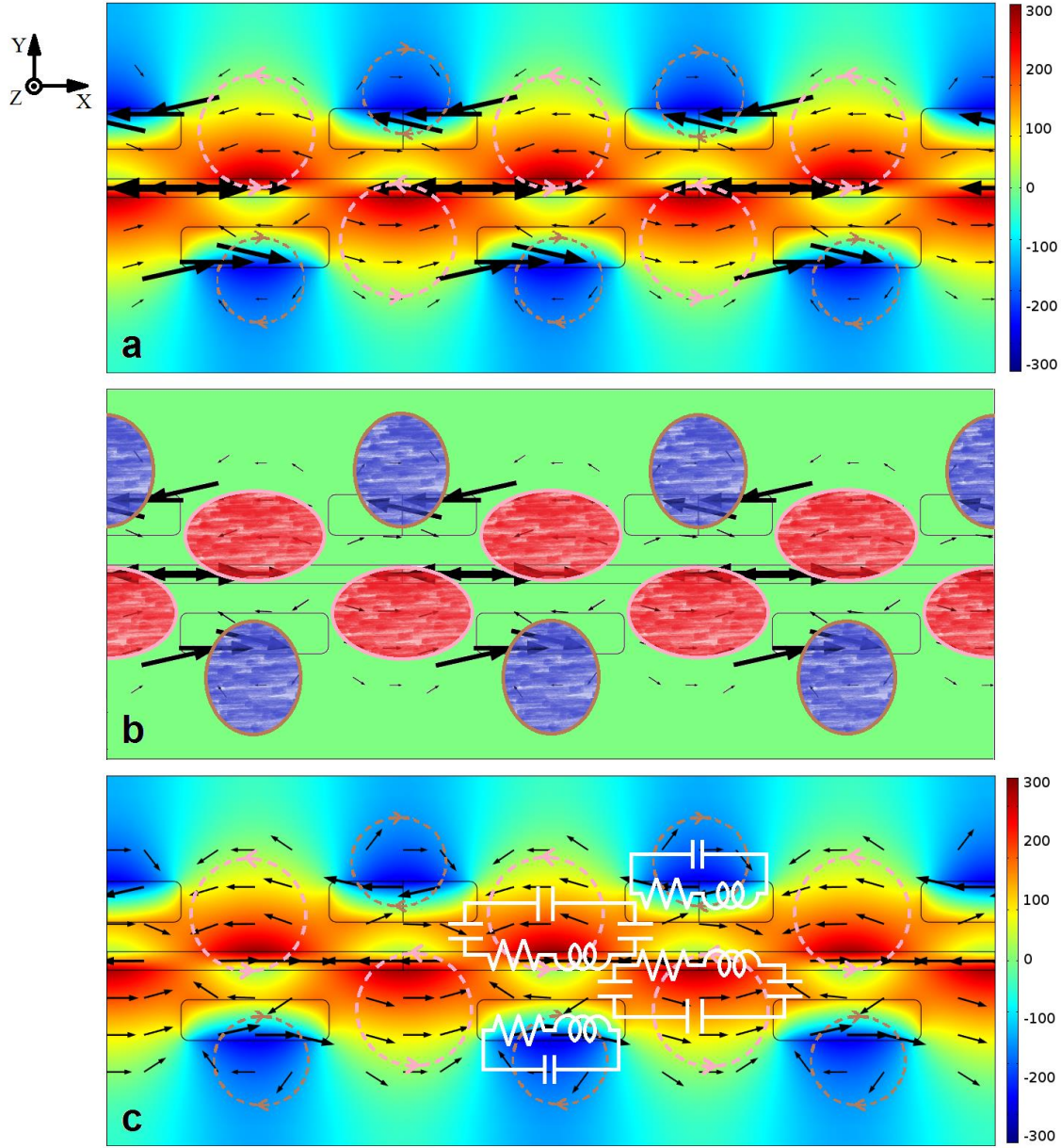

Figure S1: **Magnetic resonance mode of the negative index metamaterial underlying the proposed loss compensation mechanism:** The surface plot corresponds to the *z*-component of the magnetic field at the magnetic resonance frequency of 594THz. Finite periods of the infinite structure along the horizontal direction were shown. Grating structures have 40 nm width and 10 nm thickness and are separated from the 5 nm-thick gold film by 8 nm. (**a**) Arrows show the magnitude and direction of the induced current density. The circles indicate the loop currents forming the magnetic dipoles. Clockwise current loops are shown by brown circles and counter clockwise loops are shown by pink circles. (**b**) Simplified sketch of the magnetic fields due to approximated current loops. (**c**) Arrows indicate the direction and the magnitude of induced current density in logarithmic scale. The equivalent RLC circuits corresponding to the current loops are shown in white.

Fig. S2 shows the magnetic field distribution for the single unit cell of the NIM in Fig. S1 obtained from the eigenmode simulations using the finite integration method based commercial software package CST. This eigenmode does not only have about the same frequency as the magnetic resonance mode in Fig. S1, but also has almost the same mode profile (compare the field distribution in Fig. S2 with Fig. S1). Additionally, both modes correspond to the same propagation constant of about $1.3 \times 10^7$ rad/m. This suggests that the magnetic resonance mode excited by normally incident light in Fig. S1 is actually the eigenmode of the NIM structure.

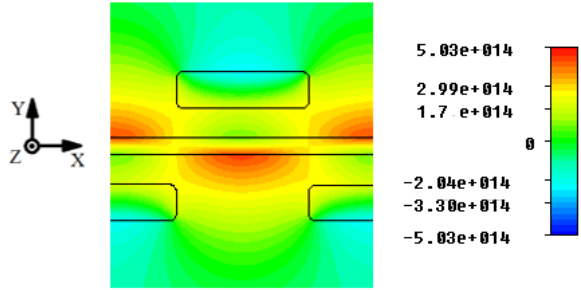

Figure S2: **Eigenmode profile of the negative index metamaterial:** The surface plot corresponds to the $z$-component (out of page) of the magnetic field (A/m). The propagation constant of the mode is approximately $1.3 \times 10^7$ rad/m and the calculated eigenfrequency is about 600THz. The paramaters are the same as in Fig. S1.

Then the critical question is, "can we coherently excite this eigenmode through multiple channels?" Owing to the unique feature of our surface plasmon driven NIM, we show below that this is indeed possible and leads to an interesting loss compensation mechanism analogous to optical gain but without requiring an optically active medium that provides gain.

Having said this, in the following we will first show how the coherent excitation of the same eigenmode, simultaneously through multiple injection channels, can be achieved using our NIM. Then, using an analytical model we will explain how such a phenomenon leads to loss compensation in the NIM without gain medium.

## 1.2. Coherent Excitation of the Eigenmode by Multiple Plasmon Injection Channels

Fig. 1 in the main text shows the magnetic field distribution inside the designed loss compensated structure at the magnetic resonance frequency when the structure is excited by side ports $P_3$ and $P_4$ only and the phase of the auxiliary fields are set equal. We observe that the mode in Fig. 1, which is excited by plasmon injection through butt coupling using the side ports is the same as the eigenmode we calculated in Fig. S2, which is at the same time the magnetic resonance mode excited by grating coupling in Fig. S1. We should note that in order to excite this mode with butt coupling both auxiliary sources are needed and this results in the excitation of a standing wave consisting of two counter propagating surface plasmon polariton (SPP) modes. Since the mode responsible from negative index in Fig. S1 is excited under normal incidence, imposing the conservation of momentum dictates

$$k_{0\parallel} = qk_g \pm k_p = 0, \tag{S1}$$

where the parallel component of the free space wave vector $k_{0\parallel} = 0$ for normal incidence, $k_g = 2\pi/\Lambda$ is the reciprocal lattice vector for the grating with period $\Lambda$, $\pm k_p$ are the wave numbers for the two counter propagating SPP modes interacting with the harmonics denoted by integers $q = \mp 1$, respectively. Therefore, in order to excite the same standing wave mode, which constitutes the eigenmode of the NIM, we need plasmon injection from both auxiliary ports. If the symmetry is broken the eigenmode of the NIM will not be constructed and the performance of the system will deteriorate.

In Fig. S3 we illustrate the compensation of losses and amplification of the transmitted signal at $n' = -1$. Fig. S3a shows wave propagation through finite NIM, defined by the central superlattice, without loss compensation while Fig. S3b is obtained under loss compensation using auxiliary sources with $\alpha = 6.95$, and keeping the input signal power the same as in Fig. S3a. We clearly see that the transmitted signal becomes visibly stronger at the output (see inside the boxes with dashed white border lines) as well as the reflectance.

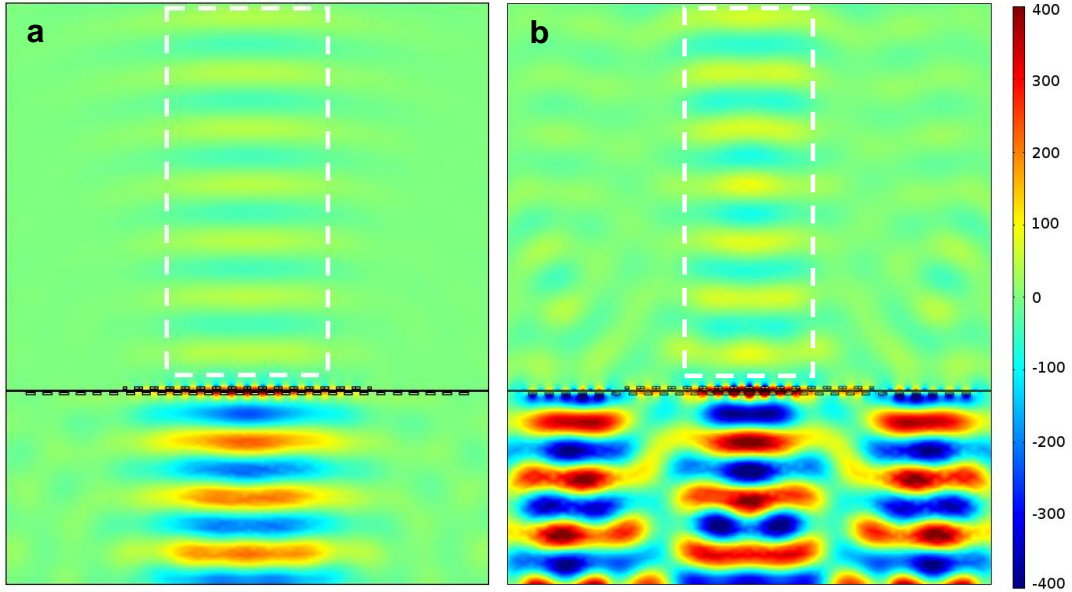

Figure S3: **Compensation of losses and amplification in the negative index metamaterial:** Wave propagation through finite period negative index metamaterial at $n' = -1$ (i.e., at 562THz) (**a**) without auxiliary sources (i.e., no loss compensation) and (**b**) with auxiliary sources (i.e., loss compensation) taking $\alpha = 6.95$. The surface plot shows the *z*-component of the magnetic field distribution. The boxed region indicates the transmitted output of the negative index metamaterial.

We also notice that the effective size of the metamaterial (or the effective aperture size) under the loss compensation is reduced. This is mainly due to the undesired interference with additional higher order modes excited by the auxiliary sources. In an ideal metamaterial, the size of the ports should not affect the transmittance. In our case this is true only in the middle part of the central superlattice as shown in Fig. S3. Therefore, the metamaterial under loss compensation is actually defined by this middle part of the central lattice and it can be possible to extend this part by appropriately engineering the high order diffraction modes, auxiliary superlattices, and using larger $\alpha$. On the other hand, the sizes of ports $P_3$ and $P_4$ mainly affect the plasmon injection rate for a given value of $\alpha$. To compensate losses in the present NIM structure, which is functional under normal incidence only (i.e., one-dimensionally functional NIM), a certain amount of power (or plasmon injection rate) should be transferred from side lattices to central lattice at a correct phase. The actual sizes of the side ports do not matter in this one-dimensional configuration as long as the ports provide the required amount of power transfer at the correct phase. There is no fixed port size to compensate the losses. If the supplied auxiliary power density is increased at the ports, the port sizes can be reduced to tune the power transfer to the required level, and vice versa. For example, for the present system, we chose the sizes of the side ports as 5 periods and $\alpha$ close to 7 to achieve the required amount of power transfer from side ports to central lattice for the compensation of losses. If we would increase the size of the side ports to 6 periods, the required amount of power transfer for the compensation of losses would be achieved at $\alpha = 7.4$ due to the reduced power density. In short, the results actually depend on the amount of transferred power and the phase necessary for the desired loss compensation not directly on the size of the ports. Therefore, it is reasonable to obtain different transmittance for different sizes of ports $P_3$ and $P_4$.

Fig. S4 shows transmittance $|S_{21}|^2$ versus phase difference $\Delta\varphi$ between the side ports ($P_3$ and $P_4$) and the input port $P_1$ for different auxiliary power levels (i.e., for different $\alpha$) at the magnetic resonance. The input fields at the side ports $P_3$ and $P_4$ are always assumed to be in phase. We notice that when there is no phase difference between the side ports and the input port $P_1$ the peak power is $0.3\pi$ radians offset from $\Delta\varphi = 0$. This offset is the result of additional path length taken by the plasmons injected from the side ports. Therefore, the perfect constructive interference between the modes excited by different channels actually occurs at $\Delta\varphi = 0.3\pi$. We can redefine the phase shift as $\Delta\varphi_o = \Delta\varphi - 0.3\pi$, so that the perfect constructive interference takes place at $\Delta\varphi_o = 0$. It is clearly seen from Fig. S4 that when the phase shift $\Delta\varphi_o = \pi$, the modes are out of phase, hence destructive interference is obtained. This is consistent with our claim that the same eigenmode (see Fig. S2) is excited coherently by both normally incident input light via port $P_1$ (see Fig. S1) and the plasmon injection via side ports $P_3$ and $P_4$ (see Fig. 1).

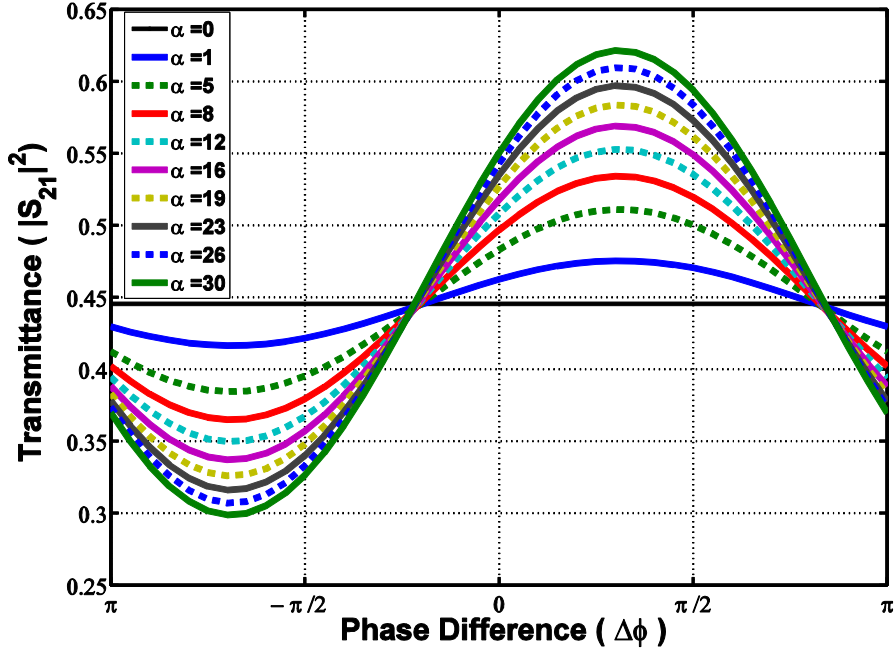

Figure S4: **Transmittance $|S_{21}|^2$ versus phase difference $\Delta\varphi$ between the side ports and the input port for different values of $\alpha$.** The eigenmode of the metamaterial in Fig. S2 is coherently excited by the side ports and the input port $P_1$. The black solid line references the transmittance under no loss compensation.

We should note that the slight variations in geometry (such as filleting, number of periods, etc.) among the simulated structures are insignificant for the purpose of explaining the underlying mechanism of the loss compensation. Rather, this indicates the robustness of the structure under loss compensation.

The commonly used figure of merit (FOM) to quantify the loss compensation in NIMs is defined as $\text{FOM} = -n'/n''$ [5-7] which we referred to as FOM$_{-1}$ at $n' = -1$ in the main text. Maximizing the FOM$_{-1}$ is particularly important for Pendry's perfect lens [8], although this imposes more stringent constraints to the loss compensation problem than maximizing the FOM at an arbitrary value of the refractive index. In the main text, we used FOM$_{-1}$ to show that the proposed mechanism of coherently exciting the same mode using multiple channels leads to the compensation of losses in surface plasmon driven NIMs (see Figs. 2 and 4). Next, we will provide an analytical description for this phenomenon.

### 1.3. Analytical Model

Consider a metamaterial consisting of periodic arrangement of SRRs similar to those multiple-gap SRRs [3, 4] prevailed in the surface plasmon driven NIM in Fig. S1 (see the current loops indicated by dashed circles defining the three-gap SRRs). Assuming the excitation of the SRRs with harmonically varying magnetic field $H_z(t) = H_z \exp(-i\omega t) + c.c.$, the magnetic flux $\phi(t)$ can be written as $\phi(t) = \mu_0 H_z(t) A$, where $A$ is the area of the loop. On the other hand, using Kirchhoff's voltage law, we can write the equation of motion for the induced electric current $I(t) = I \exp(-i\omega t) + c.c.$, on the SRR as

$$\frac{1}{C}\int I(t)dt + RI(t) + L\frac{dI(t)}{dt} = U_{ind}(t), \tag{S2}$$

where $R$, $L$, and $C$ are the effective resistance, inductance and capacitance of the simple circuit corresponding to the SRR. $I(t)$ can be evaluated from Eq. (S2) using Faraday's law of induction $U_{ind}(t) = -\partial \phi(t)/\partial t$. Once $I(t)$ is evaluated, under the quasi-static limit, the magnetization due to the magnetic dipoles can be found from

$$M_z(t) = I(t)\frac{A}{V}, \tag{S3}$$

where $V$ is the volume of the metamaterial comprising the individual magnetic dipole. Substituting $I(t)$ into Eq. (S3) after solving Eq. (S2) gives [9]

$$M_z = \chi_m(\omega)H_z. \tag{S4}$$

The effective magnetic susceptibility,

$$\chi_m(\omega) = \frac{f\omega^2}{\omega_0^2 - \omega^2 - i\gamma\omega}, \tag{S5}$$

where $f$ is the filling ratio, $\omega_0$ is the angular magnetic resonance frequency, and $\gamma$ is the damping.

When there is no loss compensation, $I(t)$ in Eq. (S3) is solely induced by external magnetic field $H_z(t)$. However, if we coherently excite the same eigenmode through multiple injection channels similar to the proposed loss compensation scheme here, $I(t)$ in Eq. (S3) should be replaced with total electric current $I_t(t) = I(t) + I_c(t)$. Let $I_c(t) = I_c \exp[-i(\omega t + \Delta\varphi_o)] + c.c.$, where $I_c(t)$ is the contribution due to the external injection.

Using Eq. (S4), the magnetization due to $I(t)$ alone can be written as

$$M_z \exp(-i\omega t) + c.c. = \chi_m(\omega)H_z(-i\omega t) + c.c. \tag{S6}$$

On the other hand, the contribution of $I_c(t)$ to the magnetization can be written as

$$M_{z,c} \exp[-i(\omega t + \Delta\varphi_o)] + c.c. = \chi_m(\omega)H_{z,c}[-i(\omega t + \Delta\varphi_o)] + c.c. \tag{S7}$$

Adding Eqs. (S6) and (S7) and assuming for simplicity $M_{z,c} = \varsigma M_z$, where $\varsigma \in \mathbb{R}$ is a scaling factor with respect to $M_z$, we obtain

$$M_z[1 + \varsigma \exp(-i\Delta\varphi_o)]\exp(-i\omega t) = \chi_m(\omega)[1 + \varsigma \exp(-i\Delta\varphi_o)]H_z \exp(-i\omega t), \tag{S8}$$

which can be simplified to

$$M_{z,t} = \chi_{m,t}(\omega)H_z, \tag{S9}$$

where $M_{z,t} = M_z[1 + \varsigma \exp(-i\Delta\varphi_o)]$ is the complex amplitude of the total magnetization and $\chi_{m,t}(\omega) = \chi_m(\omega)[1 + \varsigma \exp(-i\Delta\varphi_o)]$ is the new effective magnetic susceptibility under the loss compensation scheme. If we compare Eq. (S9) with Eq. (S4) we see that despite the same excitation field $H_z$ it may be possible to obtain larger magnetization under loss compensation due to a stronger magnetic susceptibility. For example, taking $\varsigma = 1$ and $\Delta\varphi_o = 0$ (i.e., the mode excited by auxiliary ports has the same amplitude and phase as the mode excited by the input signal field, hence constructive interference occurs) the magnetic susceptibility becomes two times larger under the loss compensation. In contrast, if we take $\varsigma = 1$ and $\Delta\varphi_o = \pi$, the magnetization vanishes due to destructive interference. These are consistent with our observation in Fig. S4. We should note that the scaling factor $\varsigma$ here has relation to $\alpha$ in the main text. The auxiliary power provided through both side ports $P_3$ and $P_4$ is assumed to be $\alpha/2$ times the input power provided through port $P_1$ (see Fig. 1). If $\alpha$ is increased, power coupled to central superlattice will be also increased, hence the scaling factor $\varsigma$, since $\varsigma$ determines the amplitude of additional magnetization $M_{z,c}$ due to the mode excited coherently through auxiliary ports. In analogy with optical gain, the function of $\alpha$ is similar to pump rate [5] or pump field amplitude [6].

Finally, based on previously defined FOM-1, we will show that the capability to modify $\chi_{m,t}(\omega)$ of the metamaterial using the method described above allows for compensating the losses in metamaterials.

In Fig. S5 we plot the impedance $z(\omega)$ and $n(\omega)$ for such an NIM with and without loss compensation. All the parameters are given in the figure caption. The imaginary part of $z(\omega)$ determines the amount of phase difference between magnetic and electric fields. Depending on the sign of the imaginary part, the electric field either leads or lags the magnetic field. On the other hand, the real part of the impedance along with the amplitude of the field determines how much power is transferred by an electromagnetic wave to the medium. Since the power transferred by an electromagnetic wave is non-negative, the real part of impedance is a positive number [10-12]. The NIM without loss compensation has FOM-1 = 1.8. However, the NIM with loss compensation has FOM-1 = 146.6. This is achieved by increasing the auxiliary field amplitude such that $\varsigma = 1.3$ and $\Delta\varphi_o = 1.7\pi$ (i.e., close to perfectly constructing interference). The NIM becomes completely loss free when $\varsigma$ is slightly larger than 1.3 (i.e., compare to Fig. 4 where the surface plasmon driven NIM becomes loss free when $\alpha$ is slightly larger than 7). If the auxiliary field amplitude is increased further, the losses are overcompensated and the material becomes an amplifying medium (i.e., $n'' < 0$).

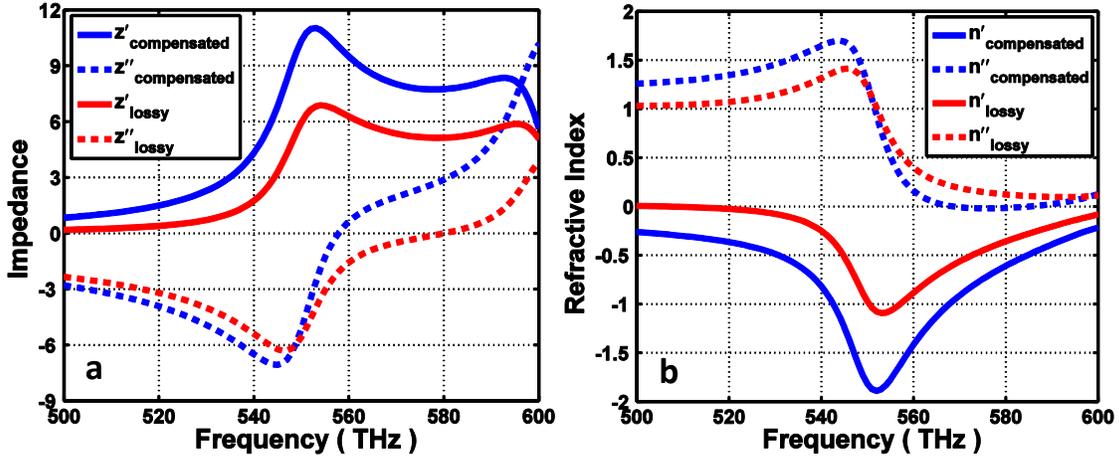

Figure S5: **The effect of the loss compensation scheme on optical parameters:** (**a**) Impedance and (**b**) refractive index for a NIM with and without loss compensation. Scaling factor $m = 1.3$, phase shift $\Delta\varphi_o = 1.7\pi$, filling ratio $f = 0.3$, angular magnetic resonance frequency $\omega_0 = 2\pi\times550$THz, damping $\gamma = 2\pi\times13$THz, angular plasma frequency $\omega_p = 2\pi\times600$THz.

### 1.4. Perfect Lens

Here, we would like to discuss briefly how our loss compensation scheme may be generalized to practical perfect lens. Our NIM structure can be an impedance matched by tuning the geometric parameters, phase, and amplitude of the auxiliary sources. For generality, however, this must be extended to isotropic NIMs. One possible way to achieve that may be starting with a passive isotropic NIM design and using a light source which is a coherent superposition of two sources, say A and B, such that the amplitude of the source B is $\alpha$ times the amplitude of source A and the phase difference between the sources is $\Delta\varphi$. Both $\alpha$ and $\Delta\varphi$ may be functions of spatial coordinates. Here, source A is the object to be imaged and B acts as the auxiliary source (see Fig. S6 for the setup). This is similar to the system in the main text except that the object to be imaged (i.e., A) and the auxiliary source (i.e., B) are supplied through the same port in the form of a coherent superposition.

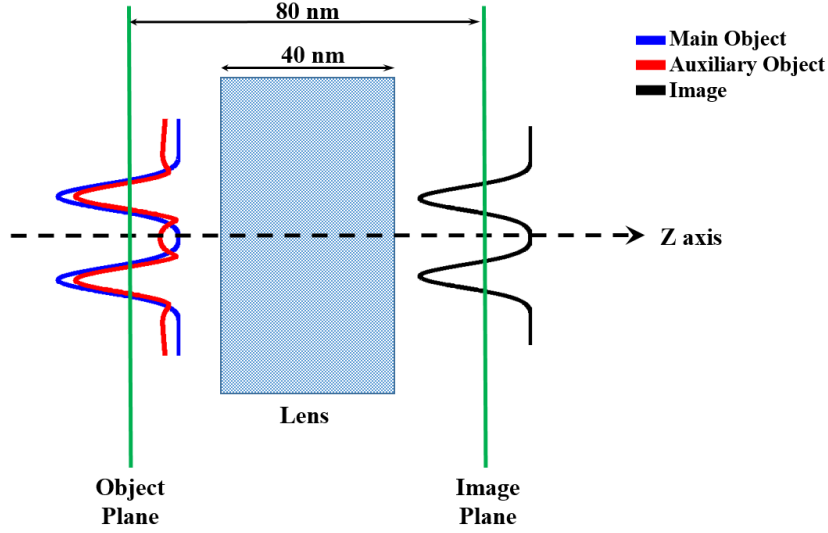

Figure S6: **Setup for the Π-scheme assisted imaging**: Coherent superposition of the main object to be imaged (i.e., referred to as A in the text) and the auxiliary object (i.e., referred to as B in the text) in the object plane results in a loss compensated image in the image plane. The thickness of the lens in the present example is 40nm. The distance between the object and image planes is 80nm.

Consider the image of a subwavelength object focused by a lossy lens (i.e., non-ideal perfect lens under electrostatic limit or poor-man's superlens) studied originally by Pendry [8]. This image can be expressed in terms of Fourier components of the source as

$$V_f(x, z = 2d) = \sum_{k_x} v_{k_x} \frac{\exp(-2k_x d)}{0.04 + \exp(-2k_x d)} \exp(ik_x x), \tag{S10}$$

where $z = 2d$ corresponds to image plane and $k_x$ denotes the spatial frequencies of the object. Note that if the constant number 0.04 would not exist in the denominator one would perfectly reconstruct the image of the source. This constant number comes from the finite imaginary part of complex permittivity (i.e., losses) used by Pendry. The question is "can we reduce or even remove this constant from the equation?"

Consider that the object to be imaged is represented by $A(x)$ and the auxiliary source by $B(x)$ at the object plane, such that $B(x) = \alpha(x)A(x)\exp[i\Delta\varphi(x)]$. Then the total input can be represented as $C(x) = A(x)[1 + \alpha(x)\exp[i\Delta\varphi(x)]]$ [similar to $M_{z,t}$ in Eq. (S9)]. If we express the contribution of $A(x)$ to the focused image as Eq. (S10), then the contribution of $C(x)$ can be written by replacing $v_{k_x}$ with $v_{k_x} f(k_x)$. That is

$$V_f(x, z = 2d) = \sum_{k_x} v_{k_x} \frac{f(k_x)\exp(-2k_x d)}{0.04 + \exp(-2k_x d)} \exp(ik_x x), \tag{S11}$$

which can be cast into

$$V_f(x, z = 2d) = \sum_{k_x} v_{k_x} \frac{\exp(-2k_x d)}{0.04p + \exp(-2k_x d)} \exp(ik_x x), \tag{S12}$$

where $p < 1$ is a constant. For example, if we choose $p = 1/2$ [i.e., the constant 0.04 in Eq. (S11) is reduced by a factor of 2 under the loss compensation], we obtain

$$f(k_x) = \frac{0.08 + 2\exp(-2k_x d)}{0.04 + 2\exp(-2k_x d)}. \tag{S13}$$

Fig. S7a below shows that the spatial frequency spectrum $v_{k_x} f(k_x)$ for the total input is not much altered compared to the object to be imaged. Thus, the input should be a coherent superposition of two similar objects, one of which acts as the auxiliary source as shown in Fig. S7b. The auxiliary source here automatically and coherently excites the underlying modes of the system *without any dependence of the device on the form of incident wave*. Under this condition the losses are reduced and the metamaterial becomes automatically impedance matched in accordance with Eq. (S12).

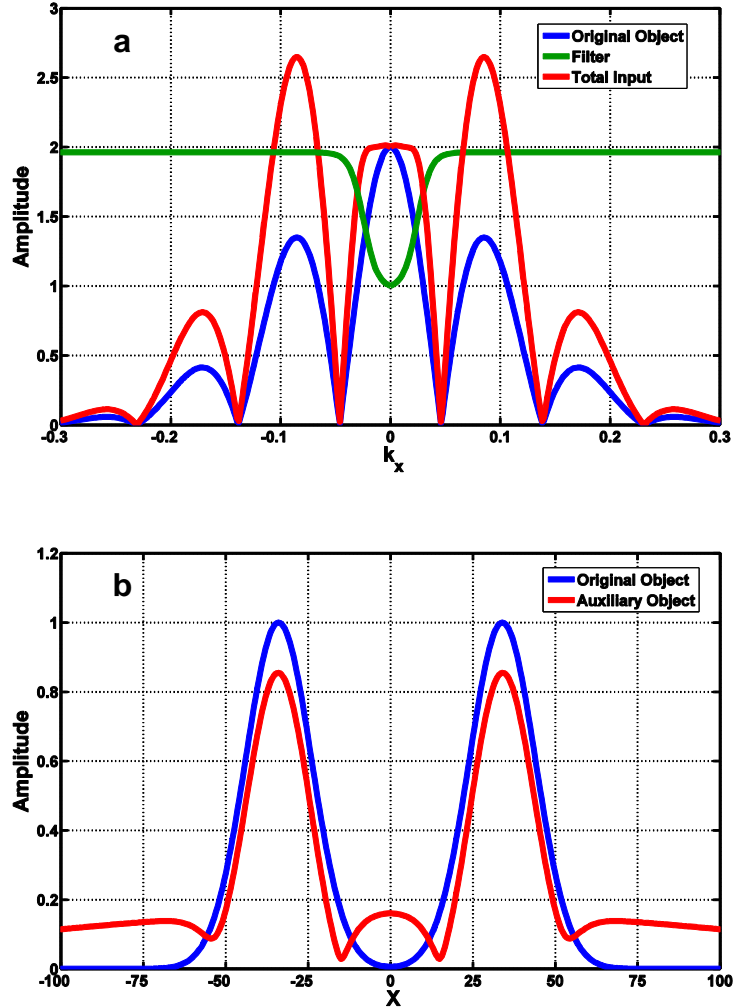

Figure S7: **Π-scheme assisted non-ideal perfect lensing.** (**a**) Spatial frequency spectrum of an object (blue) consisting of two Gaussians separated by a subwavelength distance, total input (red), and the filter function (green) given in Eq. (S13). (**b**) The amplitudes of the corresponding subwavelength object function (blue) and the auxiliary source (red) for the implementation of the Π-scheme.

One possible arrangement of the auxiliary source can be achieved as follows. Consider the original object (i.e., the actual object to be imaged, blue solid line in Fig. S7b), which we denote with $A(x)$. Assume for a moment that instead of imaging $A(x)$ we want to image an object with amplitude about half the actual object $A(x)$. Thus, half of the source acts as auxiliary source for the other half. This is equivalent to applying the filter $f(k_x) \cong 2$, which overcompensates the low-frequency Fourier components compared to the filter in Fig. S7a and is fixed in the later stage. Then $v_{k_x}$ in Eq. (S10) represents the Fourier spectrum of the total object $C(x) = A(x)$. Note that Eq. (S10) represents the image of the total object $C(x)$, which we will denote as $C'(x)$. Taking the Fourier transform of $C'(x)$ and applying the high pass filter in Eq. (S13) (i.e., green solid line in Fig. S7a), we recover the Fourier spectrum of the loss compensated image of the actual original object $A(x)$, which needs to be inverse Fourier transformed to yield the final loss compensated image of the original object $A(x)$. It is clear that the auxiliary source can be easily constructed with the help of post-processing. Importantly, since $A(x)$ is arbitrary in this process, *no a-priori knowledge of the form of the incident wave* is needed. This clearly shows the feasibility of our proposed plasmon injection method and reduces the problem to only the design of a passive isotropic NIMs.

## 1.5. Summary

In summary, under the loss compensation scheme the surface plasmons injected through side ports couple to and interfere with the negative index mode (see Figs. 1, S1-S3) supported by the central lattice. The resultant coherent superposition between injected plasmons and domestic plasmons leads to a modified susceptibility given in Eq. (S9). This in turn leads to modified impedance and refractive index such as in Fig. S5. If the phase and amplitude of the injected plasmons are appropriately selected, $n''$ may even be negative at some frequency regions. For example, in Fig. S5, $n''$ is negative between 570THz-585THz. Around such frequency region the metamaterial (i.e., central lattice) effectively behaves as an amplifying medium with a negative effective absorption coefficient as discussed in detail in, for example, Ref. [6]. In this case, the total field outgoing (i.e., reflectance plus transmittance) from the metamaterial is amplified as a result of "*the growing amplitude (i.e., amplification) of the SPPs in the central lattice owing to constructive interference with injected SPPs*" in accordance with Eq. (S8). Therefore, "*amplified SPPs*" means coherent superposition of injected and domestic SPPs in the central lattice with larger overall amplitude than domestic SPPs alone. Such SPP amplification is possible even with a positive effective absorption coefficient as long as we are in the loss compensation regime. However, in this regime the metamaterial does not behave as an amplifying medium since gain does not exceed overall losses.

Finally, it is worth mentioning that the $n'$ has relatively broad minimum around -2 (see Fig. 2a). We found that increasing the dielectric thickness has potential to shift this broad minimum to $n' = -1$. Therefore, it may be possible to achieve broadband operation by geometric tailoring with or without using, for example, multi-sized strip arrays such as in broadband perfect absorbers [13, 14].

## 2. Effective Parameter Retrieval for Loss Compensated Metamaterial

### 2. 1. Definitions for the Scattering Parameters

Port $P_1$ (see Fig. 1 in the main text) is defined as the input port and port $P_2$ as the output port. Consider the case that there are incident fields with power ξ at port $P_1$ and power β at each of ports $P_3$ and $P_4$. The field, $b$, at the output port would be

$$b = \sqrt{\xi}S'_{21} + \sqrt{\beta}S'_{23} + \sqrt{\beta}S'_{24}. \tag{S14}$$

The overall transmission coefficient, $S_{21}$, can be found by dividing the field at port $P_2$ by the field at port $P_1$. If we define $\sqrt{\frac{\alpha}{2}} = \frac{\sqrt{\beta}}{\sqrt{\xi}}$, then the equation for overall transmission coefficient would be

$$S_{21} = S'_{21} + \frac{\sqrt{\beta}}{\sqrt{\xi}} S'_{23} + \frac{\sqrt{\beta}}{\sqrt{\xi}} S'_{24} = S'_{21} + \sqrt{\frac{\alpha}{2}}(S'_{23} + S'_{24}). \tag{S15}$$

With the same approach reflected field, $a$, at port $P_1$, would be

$$a = \sqrt{\xi} S'_{11} + \sqrt{\beta} S'_{13} + \sqrt{\beta} S'_{14}. \tag{S16}$$

Therefore the overall reflection coefficient, $S_{11}$, which is the reflected field at port $P_1$ divided by input field at port $P_1$ can be found as

$$S_{11} = S'_{11} + \frac{\sqrt{\beta}}{\sqrt{\xi}} S'_{13} + \frac{\sqrt{\beta}}{\sqrt{\xi}} S'_{14} = S'_{11} + \sqrt{\frac{\alpha}{2}}(S'_{13} + S'_{14}) \tag{S17}$$

## 2.2. Calculation of the Effective Parameters

A number of adjustments has been done for the process of finding refractive index from s-parameters. We have started the process with small $\alpha$ at which the losses are not fully compensated. Therefore the imaginary part of the refractive index must be positive. With this piece of information the imaginary part of the refractive index $n''$ can be found uniquely from [12]

$$n'' = \left| Im\left\{ \frac{1}{kd} \cos^{-1}\left( \frac{1 - S_{11}^2 + S_{21}^2}{2S_{21}} \right) \right\} \right|. \tag{S18}$$

Mathematical expression for the real part of the refractive index $n'$ in terms of s-parameters is given by

$$n' = \pm Re\left\{ \frac{1}{kd} \cos^{-1}\left( \frac{1 - S_{11}^2 + S_{21}^2}{2S_{21}} \right) \right\} + m\frac{\lambda}{d}. \tag{S19}$$

Here $m$ is an arbitrary integer number. Therefore infinite number of solutions with either + sign or − sign and different $m$ could be obtained from Eq. (S19). However at most one of them could be the actual value for $n'$.

In the present Letter, the multilayer approach was exploited for two purposes: (1) to resolve the ambiguity in *m* and (2) to verify that there is a physical solution for Eq. (S19) where the structure can be approximated by length independent homogeneous effective medium. To achieve this, we found different solutions of Eq. (S19) with both + and − signs for a single layer structure and the stacks of 3, 5, and 7 layers. Then, we explored the results to find a solution for the single layer structure that agrees with the solutions for the structures with 3, 5, and 7 layers. Fig. S8 shows the solutions with + sign in Eq. (S19) for $m = -3, -2, -1, 0, 1, 2, 3$. As shown in Fig. S8, *m*= 0 for the single layer structure has good agreement with solutions for the structures with 3, 5, and 7 layers. This suggests that the homogeneous effective medium approximation is applicable to the metamaterial with *m*= 0 being the correct branch for the physical solution of Eq. (S19) [15].

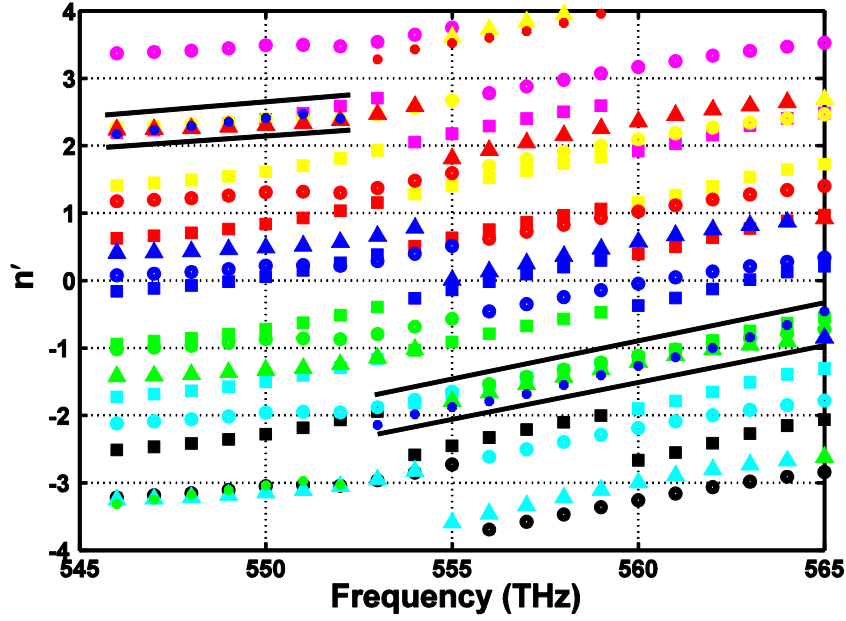

Figure S8: **Multilayer solutions:** Solutions only for the + sign in Eq. (S19) were shown for $m = -3, -2, -1, 0, 1, 2, 3$ and $\alpha = 6$. Magenta: branches for $m = 3$, Yellow: branches for $m = 2$, Red: branches for $m = 1$, Blue: branches for $m = 0$, Green: branches for $m = -1$, Cyan: branches for $m = -2$, and Black: branches for $m = -3$. ●: 1 layer, ▲: 3 layers, ○: 5 layers, □: 7 layers. The solutions confined between the black solid lines do not change with increasing number of layers.

Before we proceed further, we note that the multilayer approach we follow here does not resolve the ambiguity in the sign of Eq. (S19). The reason is as follows. Suppose that two different stacks with different number of layers $m_1$ and $m_2$ yield overlapping solutions at some frequency range $f_1 < f < f_2$ for the + sign in Eq. (S19). So we can write

$$Re\left\{\frac{1}{kd}\cos^{-1}\left(\frac{1 - S_{11}^2 + S_{21}^2}{2S_{21}}\right)\right\} + m_1 \frac{\lambda}{d} = Re\left\{\frac{1}{kd}\cos^{-1}\left(\frac{1 - S_{11}^2 + S_{21}^2}{2S_{21}}\right)\right\} + m_2 \frac{\lambda}{d}; \quad \text{(S20-a)}$$

$$for\ f_1 < f < f_2.$$

If we multiply both sides of Eq. (S20-a) by -1, we get

$$-Re\left\{\frac{1}{kd}\cos^{-1}\left(\frac{1 - S_{11}^2 + S_{21}^2}{2S_{21}}\right)\right\} - m_1 \frac{\lambda}{d} \quad \text{(S20-b)}$$
$$= -Re\left\{\frac{1}{kd}\cos^{-1}\left(\frac{1 - S_{11}^2 + S_{21}^2}{2S_{21}}\right)\right\} - m_2 \frac{\lambda}{d};$$

$$for\ \lambda_1 < \lambda < \lambda_2.$$

As it can be seen Eq. (S20-b) is also in the form of Eq. (S19), and it means solutions with $-m_1$ and $-m_2$ are overlapping for the − sign in Eq. (S19). Thus the multilayer approach here does not lead to a unique solution. Therefore, more investigation is needed to find the final solution.

In order to address the remaining ambiguity in the sign of $n'$ we investigated Kramers-Kronig (K-K) relations. As seen in Fig. S9 there are two discontinuity points in the two different solutions (i.e., + and − solutions for $m = 0$) of Eq. S19. In order to avoid such discontinuities, we considered different combinations of the + and − solutions (i..e, 8 combinations in total) as potential solutions for $n'$. Since there is a unique solution for the imaginary part, $n''$, one can apply K-K relation to find which combination gives the actual real part of the refractive index.

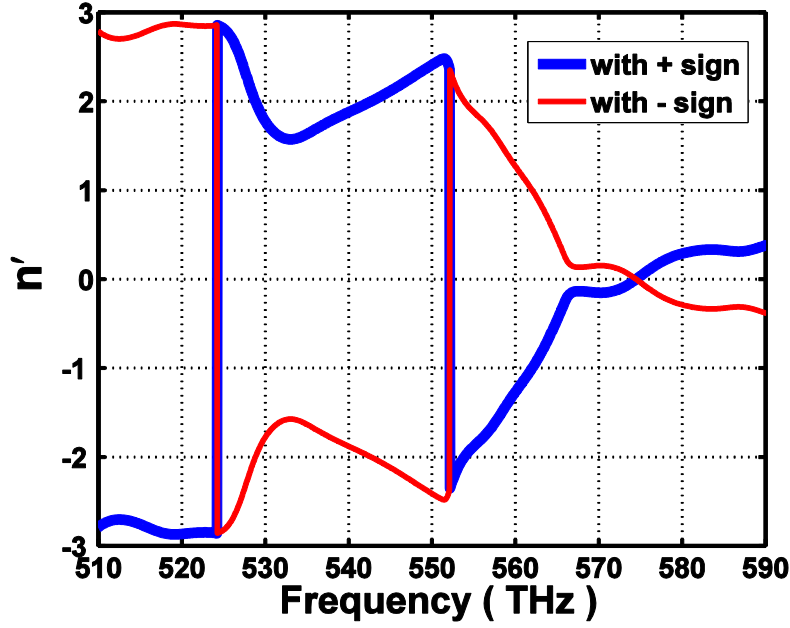

Figure S9: **Mathematical solutions for the real part of the refractive index for $m = 0$:** There exist 8 possible combinations for $m = 0$ considering both + and – signs in Eq. (S19). $\alpha = 6$.

K-K relations states $\chi'(\omega)$ and $\chi''(\omega)$ which are defined by

$$n(\omega) = \sqrt{1 + \chi'(\omega) + j\chi''(\omega)} \tag{S21}$$

satisfy the integral [16, 17]

$$\chi'(\omega) = \frac{2}{\pi} \int_0^\infty \frac{\omega' \chi''(\omega')}{\omega'^2 - \omega^2} d\omega . \tag{S22}$$

We choose each of the 8 combinations for the refractive index and find $\chi'(\omega)$ and $\chi''(\omega)$ from Eq. (S21), which are called $\chi'_{test}$ and $\chi''_{test}$, respectively. Then we plug the $\chi''_{test}$ in Eq. (S22) to obtain the corresponding $\chi'(\omega)$, which is called $\chi'_{from\,K-K}$. If $\chi'_{from\,K-K}$ agrees with $\chi'_{test}$ the corresponding solution is considered as the actual refractive index. Fig. S10 shows the closest $\chi'_{from\,K-K}$ and $\chi'_{test}$ pair we found using this routine. The corresponding refractive index is shown by the green dashed line in Fig. S11 and called $n'_{coarse}$. Although there is a discontinuity in $n'$ near 525THz, this solution has been selected because the other solutions are not consistent with the K-K relations. This solution has been refined further around the discontinuity point as explained below.

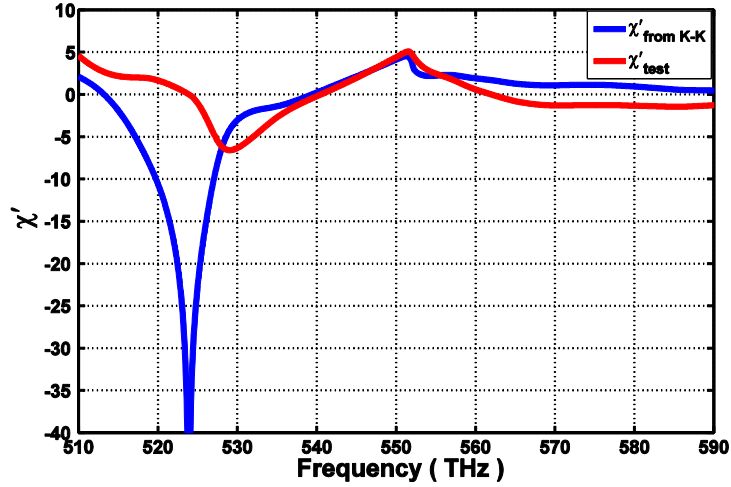

Figure S10: $\chi'$ **vs. frequency:** The closest $\chi'_{from\,K-K}$ and $\chi'_{test}$ pair obtained from Fig. S9.

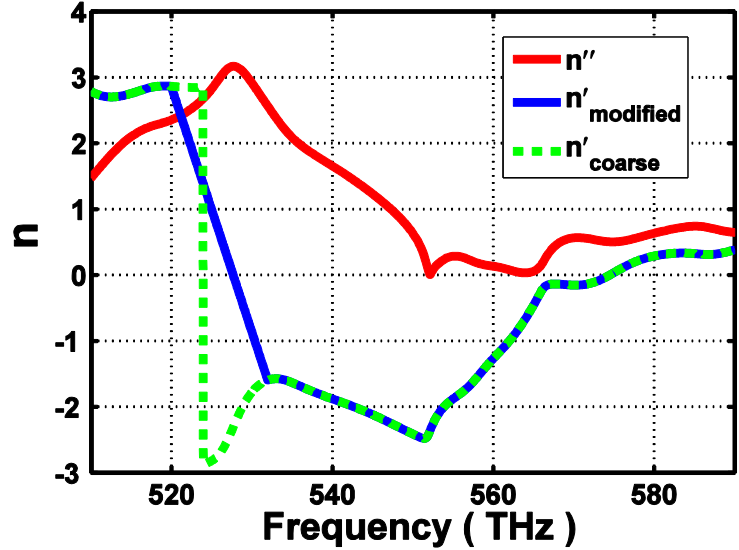

Figure S11: **Retrieved effective refractive index:** $n'$ is the retrieved real part of the refractive index obtained by the combination of multiple-layer approach and Kramers-Kronig relations. $n'_{modified}$ is the real part of the refractive index refined further around the discontinuity point near 525THz by using Kramers-Kronig relations.

As Fig. S10 shows $\chi'_{from\,K-K}$ and $\chi'_{test}$ do not agree well around 525THz. This disagreement between $\chi'_{from\,K-K}$ and $\chi'_{test}$ and the unrealistic behavior in $n'$ have the same origin and arises from numerical errors in s-parameters [11, 18]. $S_{21}$ in the denominator of Eqs. (S18) and (S19) is very small around 525THz. A minute error in numerical value of this parameter has enormous effect in the refractive index. In this frequency region (i.e., between 520THz-530THz), which is far from the region of interest, we have approximated the real part of the refractive index by a straight line using K-K relations. We have verified in Fig. S12 that the incorporation of this approximation to the retrieved refractive index leads to a good agreement between $\chi'_{from\,K-K}$ and $\chi'_{test}$ over the entire spectrum (i.e., compare Fig. S12 with Fig. S10). Fig. 2a displays this final refined refractive index, which is labeled as $n'_{modified}$ in Fig. S11. However, we removed the frequency range below 532THz due to the numerical error mentioned above. The accuracy of $n'_{modified}$ near the discontinuity point can be improved further by using higher order approximations, but this is not necessary for our purpose here.

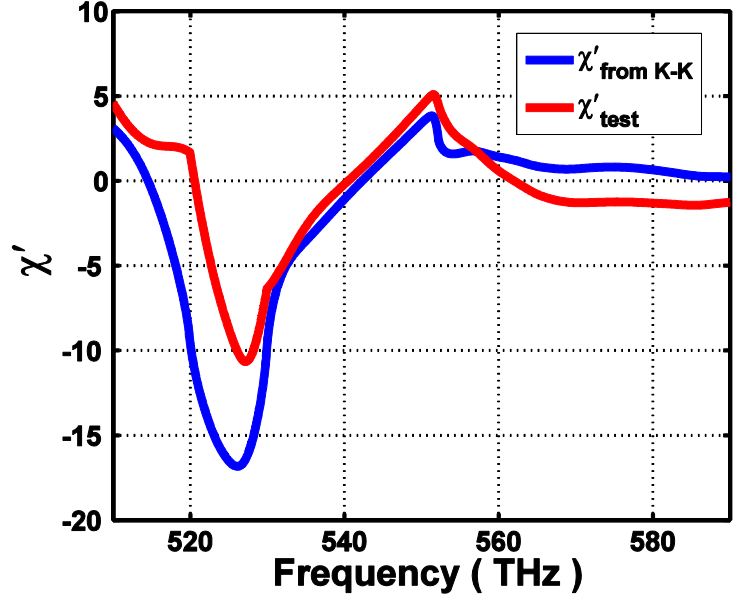

Figure S12: **Comparison of:** $\chi'_{from\ K-K}$ and $\chi'_{test}$ for $n'_{modified}$ at $\alpha = 6$.

This procedure has been repeated for different values of $\alpha$. The only difference is as we increase $\alpha$, $n''$ decreases and eventually reaches zero at some points (see Fig. S13). Considering the fact that the losses are fully compensated when $n''$ reaches zero, then the negative imaginary part means not only the losses are conpensated but also there exists some amplification. Fig. S14 shows FOM versus frequency and also indicates the frequency points corresponding to FOM-1 for different values of $\alpha$. As it is seen at $\alpha = 7$, FOM-1 starts to diverge implying that the losses are fully compensated and there is no need to increase $\alpha$ further. In addition the frequency, where $n' = -1$ occurs, changes only 0.2% as $\alpha$ increases from 0 to 7 (i.e., from no loss compensation to full compensation of losses).

Furthermore, we have verified that the wavelength inside the metamaterial structure at $n' = -1$ is approximately equal to free space wavelength and the wavefronts propagate in the backward direction. Since the thickness of the proposed structure (i.e., 100nm) is much less than the wavelength we are not able to evaluate the wavelength for the structure with a single layer. However, if we make a stack with multiple layers with a total thickness larger than the wavelength, it becomes possible to estimate the wavelength inside the structure. For the stack of seven layers, *z*-component of the magnetic field inside the structure without loss compensation is shown in Fig. S15. Although the accuracy of the wavelength estimated by this approach is limited by the finite size of the unit cells in the direction of propagation inside the multilayer stack with integer number of periods, obtained accuracy is sufficient for our purpose. From Fig. S15 and the time evolution of the field displayed in the attached animation we estimated the wavelength inside the structure somewhere between 500nm and 600nm which is close to the free space wavelength at 561THz (i.e., at $n' = -1$). Furthermore, the attached animation also shows that the wavefront moves in the backward direction consistent with negative index of refraction. To view the backward propagation in the animation clearly, consider for example the faint yellow wavefront near the upper tip of the red arrow in Fig. S15, as the time elapses this wavefront moves in the backward direction (i.e., –*y*-direction) and its amplitude increases (i.e., since the field intensity attenuates in the +*y*-direction) as it approaches to the input port. When the wavefront reaches near the bottom tip of the red arrow the original faint yellow wavefront starts to reemerge near the upper tip of the red arrow. We have verified from the time evolution of the fields that the propagation of the wavefront from the 7[th] layer to the 2[nd] layer takes slightly less than $2\pi$ change in the phase, which means then that the wavelength inside the metamaterial should be larger but close to 500nm consistent with 535nm wavelength at 561THz. A similar conclusion can be also drawn from the observation of the blue wavefronts in the animation.

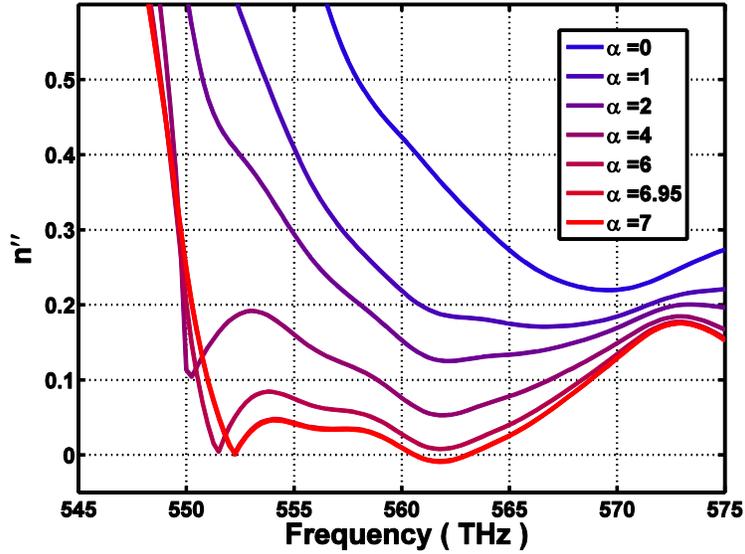

Figure S13: **Imaginary part of refractive index for different values of $\alpha$:** The imaginary part of the refractive index approaches to zero and eventually becomes negative at some frequencies with increasing $\alpha$.

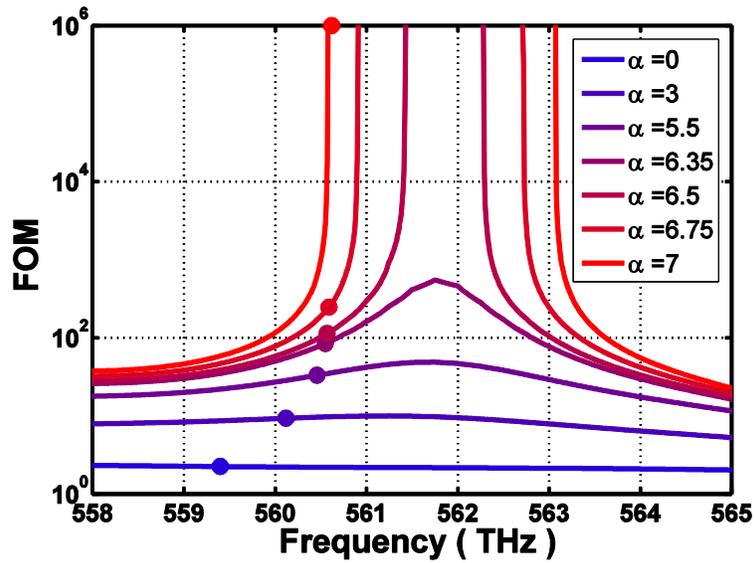

Figure S14: **FOM vs. frequency:** Solid lines: FOM, ●: FOM$_{-1}$ for different values of $\alpha$.

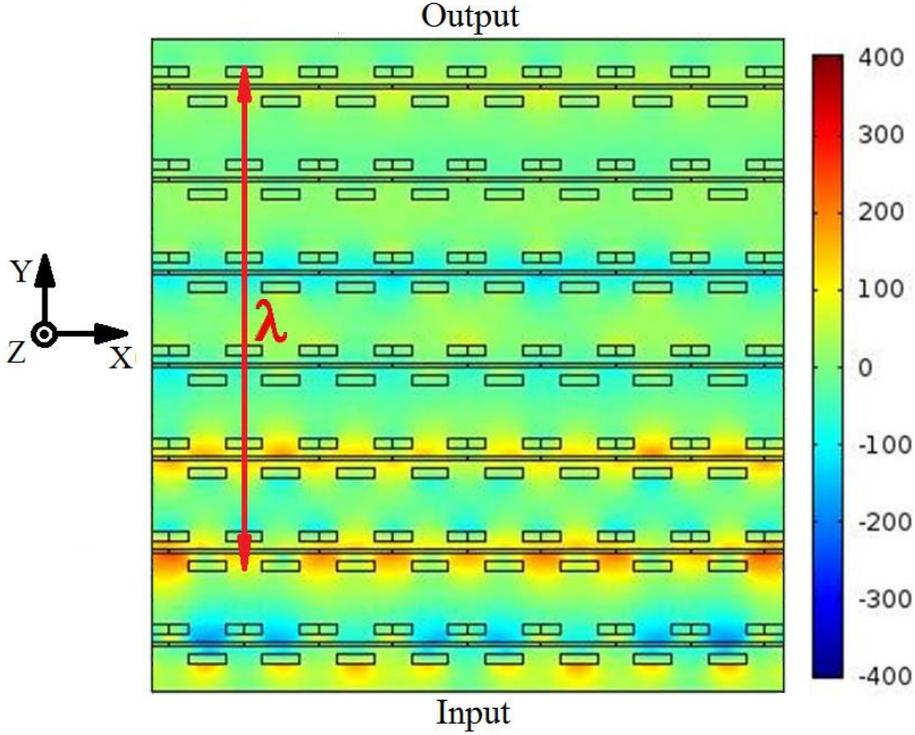

Figure S15: **Estimated wavelength inside the negative index metamaterial:** $z$- component of the magnetic field distribution at 561THz in the stack of 7 layers without loss compensation. The red arrow marks one full period. Only the central superlattice is shown.

Thus, we have verified by (1) reasonably large wavelength to slab size of larger than 5, (2) retrieved effective refractive index respecting K-K relations (see Fig. 2a), (3) length independent multiple layer retrieved effective refractive index (see Fig. S8), (4) backward propagation inside multiple layer metamaterial structure (see the animation), (5) wavelength inside the metamaterial structure estimated from field distribution (see Fig. S15 and the animation), (6) identified magnetic dipole mode (see Figs. S1 and S2) consistent with absorption spectra that the metamaterial homogenizes convincingly well.

### 3. Device Characterization

If there is no power at port $P_1$, only with power input at the auxiliary ports $P_3$ and $P_4$, some power is still delivered to the output port $P_2$. The field distribution inside the structure for this scenario is displayed in Fig. 1 of the main text. Fig. S16 shows the output power at port $P_2$ normalized to the power at the auxiliary ports (i.e., the powers at the auxiliary ports are always equal) vs. frequency when the input power at port $P_1$ is zero. Note that the percentage of the power delivered to port $P_2$ from the auxiliary ports especially around the region of interest (i.e., around 560THz, see Fig. 2a) is small.

In order to further explain the behavior of the structure, output power at port $P_2$ vs. input power at port $P_1$ for different total auxiliary power levels (i.e., equally distributed between ports $P_3$ and $P_4$) is shown in Fig. S17 (in arbitrary units). Let us call the output power at port $P_2$ as the leakage output power when the input power at port $P_1$ is zero. As it is seen in Fig. S17 the leakage power ranges from 0 to 0.05 as the power at the auxiliary ports increase from 0 to 7 units. If we compare the leakage power with the output power when the loss is fully compensated [i.e., this occurs when $\alpha$ is about 7, see Fig. 4, or in other words when the auxiliary power is 3.5 times the input power, see Eq. (1) and Section 2.1], we estimate from Fig. S17 that the leakage is near 15% of the output power regardless of the auxiliary power level. Thus, although the leakage power is not zero, it is much less than the output power when the loss is compensated.

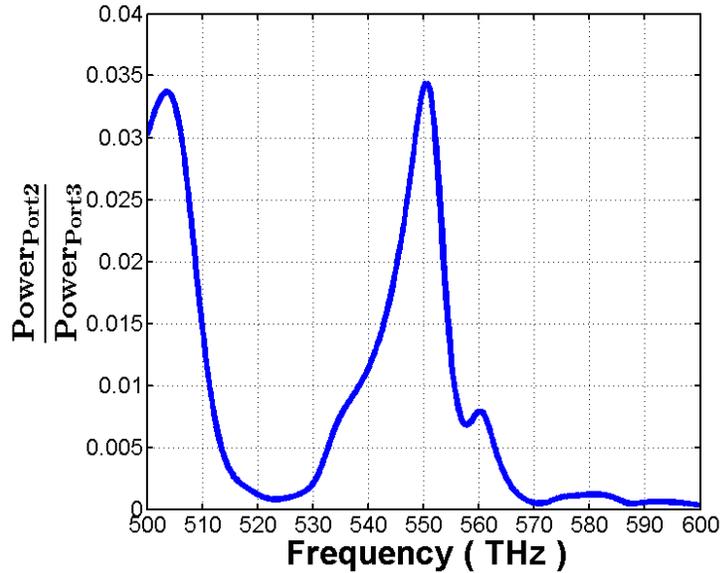

Figure S16: **Normalized output power vs. frequency**: The input power at port $P_1$ is zero. Output power at port $P_2$ is normalized to the power at the auxiliary port $P_3$, which is always assumed to have the same power as the other auxiliary port $P_4$.

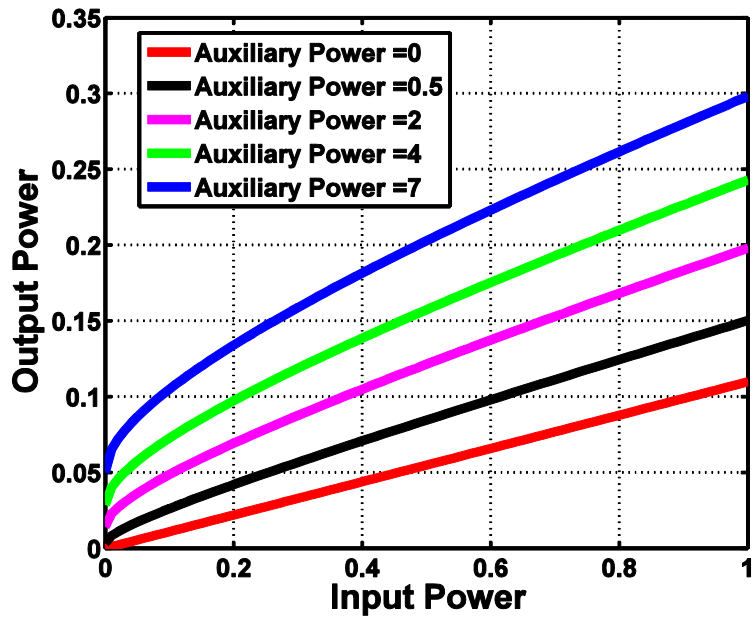

Figure S17: **Output power vs. input power:** Output power at port $P_2$ versus input power at port $P_1$ for different total auxiliary powers equally distributed between ports $P_3$ and $P_4$. Arbitrary units are used.

Furthermore, consider for example the case that input power is 1 unit and the total power at the auxiliary ports is 7 units (i.e., each auxiliary port supplies 3.5 units power). As a result, output power would be 0.3 units. When there is no input power but only auxiliary power, there is 0.05 units of output power (i.e., leakage power). If there is no auxiliary power but only 1 unit of input power at port $P_1$, the output power is about 0.11 units implying the existence of loss. When we supply both the auxiliary power (total 7 units) and the input power (1 unit) simultaneously, the total output power at port $P_2$ (i.e., 0.3 units) exceeds the sum of the powers delivered individually through the auxiliary ports and the input port (i.e., 0.16 units). We have explained the physical phenomenon behind this observation in detail in Section 1 based on coherent superposition of domestic and injected plasmons in the metamaterial.

Fig. S18 compares the output power at port $P_2$ normalized to input power at port $P_1$ vs. frequency for the structure with and without grating at two sides. No power is provided at the auxiliary ports $P_3$ and $P_4$. Under this condition both structures behave almost the same because the leakage power from port $P_1$ to port $P_3$ and $P_4$ is small, as shown in Fig. S19.

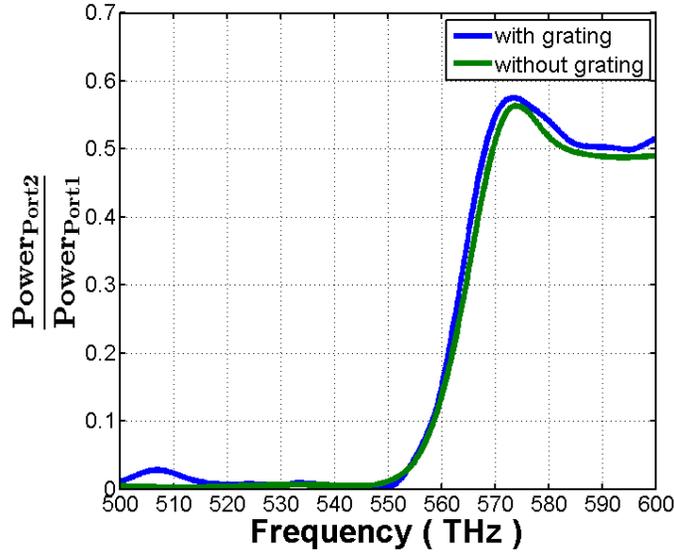

Figure S18: **Normalized output power vs. frequency:** Output power at port $P_2$ is normalized to the input power at port $P_1$. No power is provided at the auxiliary ports $P_3$ and $P_4$.

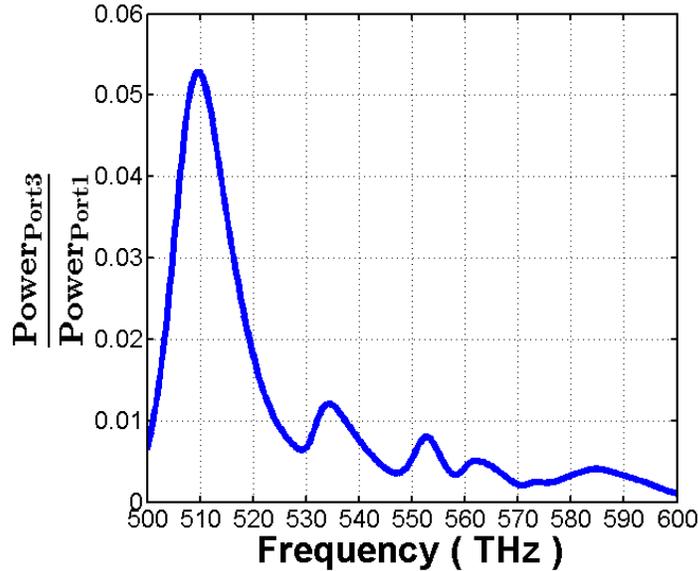

Figure S19: **Normalized power at the auxiliary port vs. frequency:** Power at the auxiliary port $P_3$ is normalized to input power at port $P_1$. The power at the auxiliary port $P_4$ is the same as $P_3$ due to symmetry.

Fig. S20 shows total transmittance with gratings at two sides (i.e., the ratio of the power delivered to output port $P_2$ to the total power supplied by the ports $P_1$, $P_3$, and $P_4$), transmittance of the structure without grating (i.e., the ratio of the power delivered to output port $P_2$ to the input port $P_1$), and the transmittance $|S_{21}|^2$ under the loss compensation defined in Eq. (1-b) in the main text. The results show that at the frequency range of 540THz-554THz the total transmittance with gratings at two sides is larger than the transmittance of the structure without gratings because the coupling (or plasmon injection rate) from side superlattices to the central superlattice is maximum at this range. In the frequency region above 554THz the transmittance without grating starts to become substantially large compared to the transmittance with gratings because the metamaterial already has a good transmittance in this frequency region compared to coupling from the side superlattice to central lattice (i.e., metamaterial). However, as it is seen in Fig. S20, despite low coupling from side superlattice to central superlattice, $|S_{21}|^2$ exceeds both the total transmittance with gratings and the transmittance without gratings for most frequency regions including the negative index region between 530THz-575THz (see Fig. 2a for the refractive index plot). It is interesting to note that above 580THz, $|S_{21}|^2$ is less than the transmittance without gratings due to destructive interference in the metamaterial between injected plasmons and domestic plasmons (see Section 1).

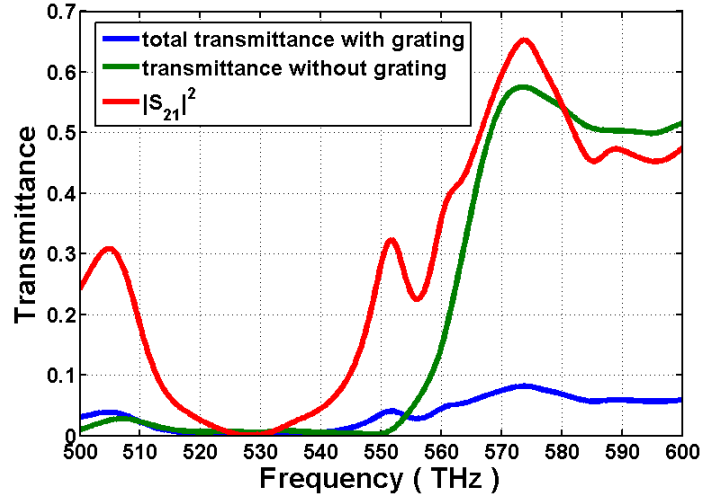

Figure S20: **Transmittance vs. frequency.** Total transmittance with gratings at two sides of the metamaterial, transmittance without gratings, and transmittance under loss compensation $|S_{21}|^2$ for $\alpha = 7$.

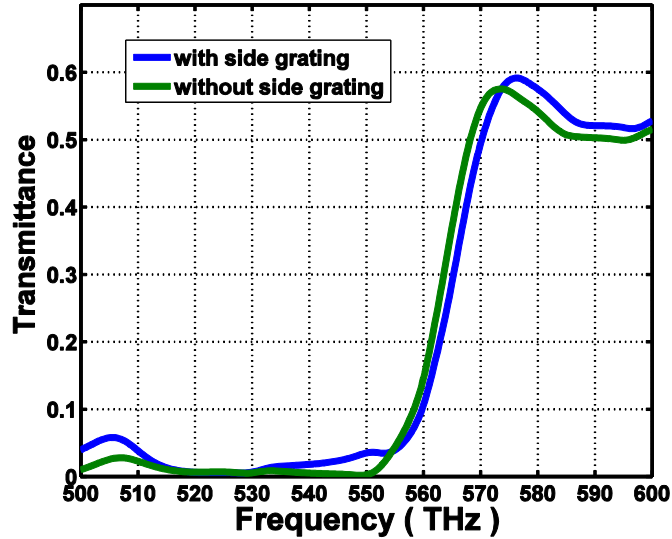

Figure S21: **Transmittance for back propagation:** The transmittance for the structures both with grating and without grating are shown.

In the case of back propagation, Fig. S21 shows the transmittance for the back propagation for the two different cases. When there is no excitation at the auxiliary ports $P_3$ and $P_4$, the structures with grating and without grating manifest very close behavior due to small leakage from central superlattice to auxiliary lattice like in Fig. S19.